\documentclass[
aps,
prl,
preprint,
superscriptaddress,
floatfix
]{revtex4-1}
\usepackage{amsmath}
\usepackage{graphicx}
\usepackage{multirow}

\begin{document}
\title{Simulating the effect of boron doping in superconducting carbon}
\author{Yuki Sakai}
\affiliation{Center for Computational Materials, Institute for Computational Engineering and Sciences, The University of Texas at Austin, Austin, Texas 78712, USA}
\author{James R.~Chelikowsky}
\affiliation{Center for Computational Materials, Institute for Computational Engineering and Sciences, The University of Texas at Austin, Austin, Texas 78712, USA}
\affiliation{Department of Chemical Engineering, The University of Texas at Austin, Austin, Texas 78712, USA}
\affiliation{Department of Physics, The University of Texas at Austin, Austin, Texas 78712, USA}
\author{Marvin L.~Cohen}
\affiliation{Department of Physics, University of California at Berkeley, Berkeley, California 94720, USA}
\affiliation{Materials Sciences Division, Lawrence Berkeley National Laboratory, Berkeley, California 94720, USA}

\date{\today}

\begin{abstract}
 We examine the effect of boron doping in superconducting forms of amorphous carbon. By judiciously optimizing 
 boron substitutional sites in simulated amorphous carbon we predict a superconducting transition temperature near 
 37~K at 14~\% boron concentration.   
 Our findings have direct implications for understanding the recently discovered high T$_c$ superconductivity in Q-carbon.  
\end{abstract}


\maketitle

Doped carbon materials are known to be superconductors with relatively low values for T$_c$.
For example, graphite intercalation compounds exhibit superconductivity \cite{Hannay1965} with
alkali and alkaline-earth metal intercalant with a superconducting transition temperature $T_c$, up to 11.5~K 
in the case of CaC$_6$ under ambient pressure \cite{Weller2005}. 
Alkali-doped fullerene (C$_{60}$) solids also exhibit superconductivity \cite{Hebard1991}, {\it e.g.},   
$T_c$ is 33K in the case of Cs$_x$Rb$_y$C$_{60}$ \cite{Tanigaki1991}.
In addition to these $sp^2$-hybridized carbon materials, boron-doped $sp^3$-hybridized 
diamond is also known to be a superconductor \cite{Ekimov2004}.
The highest $T_c$ experimentally for this material is 11.4~K \cite{Takano2007} although the theoretically 
predicted potential $T_c$ is approximately 55-80~K \cite{Moussa2008}. 

One obstacle in the realization of a high $T_c$ is the limited amount of boron that can be doped in diamond. 
Recently, a T$_c$ of 36.0~K has been reported when boron atoms are doped into a new amorphous form 
of carbon: \textit{Q-carbon} \cite{Narayan2015,Bhaumik2017}.
The amount of boron doping reported is approximately 17~\%. 
The $T_c$ of this boron-doped Q-carbon is comparable to that of MgB$_2$ (39~K) \cite{Nagamatsu2001}.  
However, the structural and electronic properties of boron-doped Q-carbon remain largely unknown 
in spite of its remarkable $T_c$.

Here we report the effect of boron doping into amorphous carbon in an attempt to understand Q-carbon and related materials.  
We replace the carbon atoms of a simulated amorphous carbon system with boron atoms one by one and
find that the acceptor states can be either shallow or deep depending on surrounding geometries. 
We show that shallow acceptor states, which are important for achieving superconductivity 
in boron-doped diamond \cite{Lee2006}, can be realized when we properly choose specific substitutional sites. 
We also study the electron-phonon coupling in boron-doped amorphous carbon
and show that those with shallow acceptor states can induce a superconductivity as in boron-doped diamond. 
We find a $T_c$ of 37~K  can be achieved at 14~at\% boron with a resulting 
electron-phonon coupling constant, $\lambda$ = 1.11. 

We employ a total energy pseudopotential approach with 
Troullier-Martins norm-conserving pseudopotentials\cite{Cohen1982,Ihm1979,Troullier1991} 
constructed within density functional theory (DFT) \cite{Hohenberg1964,Kohn1965} 
using the local density approximation \cite{Ceperley1980,Perdew1981}. 
The real-space pseudopotential DFT code PARSEC is used for molecular
dynamics (MD) simulations and structural determination 
of amorphous carbon \cite{Chelikowsky1994,Chelikowsky2000,Kronik2006,Natan2008}.  
Once the structure is determined, the plane-wave DFT package Quantum ESPRESSO \cite{Giannozzi2009} is used for phonon
and electron-phonon coupling calculations based on density functional perturbation theory \cite{Baroni2001,Giustino2017} 
with $\Gamma$-only sampling.
The McMillan equation as modified by Allen and Dynes is adopted to estimate the $T_c$ 
from $\lambda$, logarithmic average phonon frequency $\omega_{log} = \mathrm{exp}(\left<\mathrm{ln}\omega\right>)$, 
and the Coulomb repulsion parameter $\mu^* = 0.12$ \cite{Allen1975}.
See Supplemental Material for the detailed information on DFT calculations \cite{SM}.

We perform first-principles Born-Oppenheimer MD simulations
as implemented in the PARSEC code first to obtain randomized atomic coordinates of liquid-like carbon \cite{SM}. 
A cubic cell with 64 carbon atoms is employed to model the amorphous system.
We pick up different randomized atomic coordinates, {\it i.e.}, snapshots at various time steps of the above MD run
and ``quench'' those coordinates to obtain an amorphous structure.
It has been argued that a higher density coupled with a faster quenching rate generally give a larger portion of
4-fold coordinated atoms in amorphous carbon \cite{Marks1996,Mcculloch2000,Han2007}.
Based on the dominant 4-fold coordination observed in Q-carbon \cite{Narayan2015}, 
we choose a relatively high density of 3.4~g/cm$^3$ (lattice constant of the cubic cell is 7.21~{\AA}) 
that is close but lower than that of diamond to obtain 4-fold-rich amorphous carbon structure.
In addition, we immediately quench (relax) the structure from a randomized carbon atoms to 
achieve high 4-fold coordination. 

\begin{figure}[thp]
  \includegraphics[width = 8.6cm]{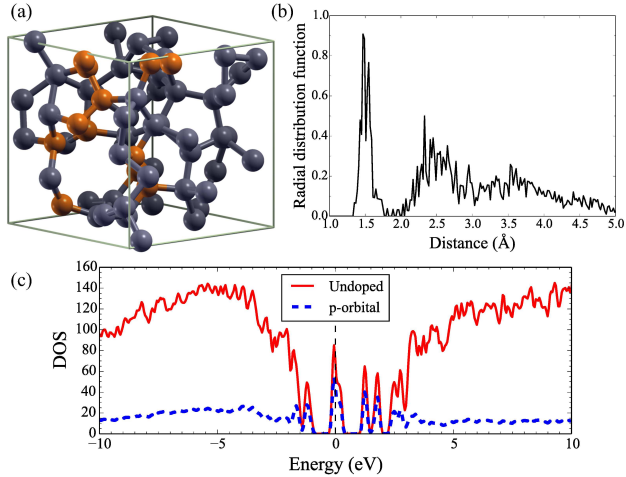}
    \caption{
    \label{fig:0dos}
(Color online) (a) Ball-and-stick model of amorphous carbon. Gray and orange spheres represent
3-fold and 4-fold coordinated carbon atoms, respectively.
(b) Radial distribution function of amorphous carbon. 
(c) Density of states (in states/spin/Ry/cell) of undoped amorphous carbon (red solid line) and 
its projection onto $p$-orbital of 3-fold coordinated carbon atoms (blue dashed line).
The vertical dashed line at 0~eV indicates the Fermi level.
A Gaussian broadening width and energy grid of 0.05~eV is used. 
}
\end{figure}

The amorphous carbon studied here is the lowest energy structure 
among several different structures generated by the above procedure (see Supplemental Material for
the atomic coordinates) \cite{SM}. 
The coexistence of the 3- and 4-fold coordinated atoms can be seen in the ball-and-stick model and
can be also recognized by the broad peak around 1.5~{\AA} in the radial 
distribution function [see Figs.~\ref{fig:0dos}(a) and \ref{fig:0dos}(b)]. 
This structure contains 81~\% (12 atoms of 64 atoms) of 4-fold-coordinated carbon atoms and rest is 3-fold 
(we use 1.8~{\AA} as the criterion for the coordination).
This value is close to the experimental observed high $sp^3$ ratio of 85~\% in Q-carbon \cite{Bhaumik2017}.

We find the density of states of the amorphous carbon 
to have several localized states near the Fermi energy as shown in Fig.~\ref{fig:0dos}(c).
These localized electronic states predominantly consist of $p$-orbitals of the 3-fold coordinated carbon atoms 
as the projected density of states (blue dashed line) indicates.
The $p_z$-orbital of 3-fold carbon atoms in amorphous carbon 
cannot form a $\pi$-network as in graphite particularly when the 3-fold atom is surrounded by 4-fold atoms. 
The proximity of 3-fold and 4-fold atoms form localized states in the gap region. 
Even when a 3-fold carbon atom is not completely isolated, the atom can be seen as a defect in a 4-fold dominated carbon system. 
Such an imperfect $sp^3$-hybridization causes the shoulder-like feature around the gap edges as well.  

\begin{figure}[thp]
  \includegraphics[width = 8.6cm]{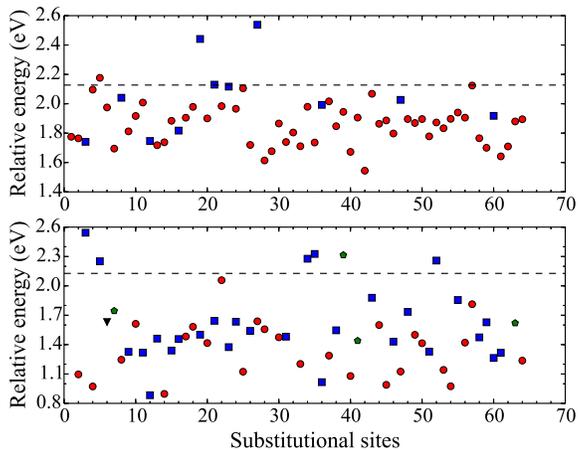}
    \caption{
    \label{fig:eig}
(Color online) Site dependence of relative Kohn-Sham eigenvalue energies of C$_{64}$ to BC$_{63}$ case (top panel) 
and B$_5$-S1 to B$_6$ case (bottom panel). 
Black triangles, blue squares, red circles, and green pentagons represent 
2-fold, 3-fold, 4-fold, and 5-fold coordination, respectively (2-fold is almost 4-fold since it has
two nearest-neighbor atoms at a distance slightly longer distance than 1.8~{\AA}. 
Here the energy difference between 128th and 123rd states are plotted. 
The horizontal dashed line shows the corresponding value in the undoped case.
}
\end{figure}

We systematically replaced carbon atom sites with boron atoms 
to investigate the effect of substitutional boron doping on electronic and superconducting properties 
\footnote{It is difficult to obtain even metallic boron-doped amorphous carbon when we 
randomize and quench atomic coordinates of B and C coexisting system.
We usually obtain deep localized impurity states, resulting in semiconductors.}.
An important feature in superconducting boron-doped diamond is hole doping, owing to
shallow acceptor states close to the occupied states \cite{Lee2006}. 
However, in amorphous carbon, and presumably Q-carbon, the position of the acceptor states strongly depends 
on the geometries of substitutional sites. For example, the relative energy of the highest occupied 
state in our simulation for the undoped case (the 128th state) measured from a reference state ({\it e.g.}, 
the 123rd state when the target doping amount is 9 or 10) is 2.12~eV at the $\Gamma$ point in the first Brillouin zone.
This relative energy of the 128th state varies from 1.54~eV in one site to 2.54~eV in a different site 
when we substitute one of the 64 carbon atoms with a boron atom. (See the top panel of Fig.~\ref{fig:eig}). 
Therefore, there is no trivial way to obtain a dominant hole-doped electronic structure. 
One needs to carefully choose substitutional sites. 

Based on this observation: 
(1) Substitutional sites are chosen so that the 128th 
state becomes low compared with the reference state (123rd state here) at the $\Gamma$ point when the lowering is significant.
(2) If the lowering is not significant, we use the $\Gamma$-point energy eigenvalues to screen possible substitutional sites. 
(3) We compute the density of states to further screen the sites in question. 
(4) The electron-phonon coupling constant of the optimized candidate structure is computed. 
This reduces the number of electron-phonon coupling calculations,  which can be computationally intensive. 

We investigate a favorable substitutional site to create shallow acceptor states from Fig.~\ref{fig:eig}.
From C$_{64}$ to BC$_{63}$, the ``best'' ten substitutional sites are all 4-fold coordinated as expected.
Eight of these best 10 sites have 4-fold coordinated nearest-neighbor atoms.
In addition, the nearest-neighbor distance is less than 1.64~{\AA} and in 6 cases, the distance is less than 1.60~{\AA}. 
On the other hand, the half of the ``worst'' ten sites are 3-fold coordinated.
The rest is 4-fold-coordinated, but four of the five remainder has a long nearest neighbor distance above 1.7~{\AA}. 
The other one has two 3-fold coordinated atoms in its nearest neighbor.
This result indicates 3-fold coordinated sites are generally unfavorable to create shallow acceptors.

The distribution of eigenvalues is also analyzed in one of the cases from B$_5$C$_{59}$ to B$_6$C$_{58}$ to
assess the interaction with preexisting boron atoms (the bottom panel of Fig.~\ref{fig:eig}). 
Seven of the worst ten sites are 3-fold coordinated, 
and two 4-fold coordinated sites turn a preexisting 4-fold B atom into 3-fold atom. 
On the other hand, eight of the best ten sites are 4-fold coordinated and preexisting boron atoms remain 4-fold
in most cases although a few are 5-fold coordinated.  
The other two sites have a boron atom at 1.8~{\AA} and 2.0~{\AA}, and can be virtually regarded as 4-fold coordinated
considering the size of boron atom.
In addition, a 3-fold coordinated boron-dimer creates a deep acceptor state in the 5th site 
while a 4-fold boron dimer in the 47th site creates a shallow state. 
Therefore, the creation of 3-fold coordination is again not favored and 4-fold coordination is favored
even when boron dimers are formed.
\footnote{Boron substitution also modifies the geometry of surrounding atoms 
because the amorphous structure has more room for relaxation than crystal. 
This significantly modifies the electronic properties as well, and 
it is still difficult to predict a favorable site to create shallow acceptor beforehand.}.

\begin{figure}[tbp]
  \includegraphics[width = 8.6cm]{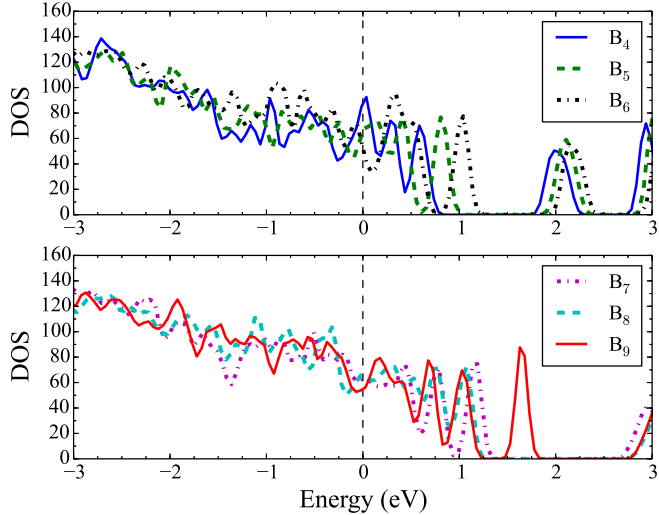}
    \caption{
    \label{fig:dos}
(Color online) Densities of states (DOS, in states/spin/Ry/cell) of B$_4$-S1 (blue solid), 
B$_5$-S1 (green dashed), B$_6$-S1 (black dash-dotted), B$_7$-S1 (magenta dash-dotted), B$_8$-S1 (cyan dashed), 
and B$_9$-S1 (red solid) structures. 
}
\end{figure}

Figure~\ref{fig:dos} illustrates the evolution of the densities of states of boron-doped amorphous carbon
starting from B$_4$-S1 to B$_9$-S1 structures following the procedure explained above.
Here S$x$ represent a series of doped structures.
In principle, the Fermi level becomes deep compared with the band edge as we increase the number of boron atoms.
The exception is the case from B$_7$S1 to B$_8$S1 where the state at the relative energy of the band-edge 
state around 1~eV becomes shallower. 
In the B$_9$-S1 case, the 128th state become deeper (above 1.5~eV) 
because the 59th site is 3-fold coordinated and it creates a relatively localized state as will be discussed below. 
Consequently, we are able to obtain shallow acceptor states in boron-doped amorphous carbon by properly selected substitutional doping.

The total energies of boron-substituted amorphous carbon also depend on the substitutional site. 
The relaxation of atomic coordinates caused by substitution affects most of the atoms in the unit cell, and
the energy variation can be relatively high compared with the energy scale of the case of the dopant in crystal.
Nevertheless, the energy difference between the lowest and highest energy structures is 53~meV/atom in
the case from C$_{64}$ to BC$_{63}$.
This implies that any doping situation can be realized in an extremely 
high-temperature and fast-quenching synthesis of boron-doped amorphous carbon. 
Therefore, we may choose ideal substitutional sites to obtain shallow acceptor states. 

Table~\ref{tb} lists the structure, $\lambda$, 
square and logarithmic average frequency $\omega_{log}$ and $\omega_2 = \sqrt{\left<\omega^2\right>}$, 
$T_c$, and Fermi level density of states 
$N_{E_F}$ of various boron-doped amorphous carbon.
In principle, $\lambda$ and $T_c$ increases as the number of dopants increases while the $\omega_{log}$ decreases.
Starting from the B$_4$-S1 case ($T_c$ of 15~K), the $T_c$ becomes 19, 23, 26, and 33~K.
The highest $T_c$ of 37~K is achieved in the 14~\% doping case, which is
comparable to the experimentally measured $T_c$ of boron-doped Q-carbon.\cite{Bhaumik2017} 
Table~\ref{tb} also lists another series of boron-doped amorphous carbon
starting from B$_4$-S2 structure to show that another doped structure exhibits electron-phonon coupling as well. 
\begin{table}
\caption{\label{tb}
Electron-phonon coupling constant ($\lambda$), logarithmic average frequency ($\omega_{log}$), 
square average frequency ($\omega_2$), superconducting transition temperature $T_c$, 
and density of states at the Fermi energy ($N_{E_F}$, in states/spin/Ry/cell) of various boron-doped amorphous carbon systems. 
A branch in series S$x$ is represented with an alphabet suffix as S$x_a$.
The values in the parentheses are those without the contribution from the lowest-frequency vibrational mode.
The convergence check of these parameters with respect to the Gaussian broadening 
width and atomic coordinates of the structures are provided in Supplemental Material \cite{SM}.
}
\begin{ruledtabular}
    \begin{tabular}{cccccc}
         Structure  & $\lambda$  & $\omega_{log}$ (K) & $\omega_2$ (K) & $T_c$ (K) &  $N_{E_F}$\\
        \hline                                                         
        \multicolumn{6}{c}{B$_4$ (6.2~\%)}  \\                           
            S1      & 0.64 (0.41)& 593 (839)          &  850 (1034)    &  13 (2)   &  69.0     \\
            S2      & 0.47 (0.44)& 807 (840)          & 1029 (1053)    &   5 (4)   &  71.3     \\
        \hline                                                         
        \multicolumn{6}{c}{B$_5$ (7.8~\%)}  \\                           
            S1      & 0.67 (0.49)& 603 (770)          &  865  (997)    &  16 (6)   &  61.1     \\
            S2      & 0.56 (0.50)& 769 (852)          & 1003 (1058)    &  11 (7)   &  57.8     \\
        \hline                                                         
        \multicolumn{6}{c}{B$_6$ (9.4~\%)}  \\                           
            S1      & 0.69 (0.58)& 672 (807)          &  920 (1000)    &  19 (13)  &  50.8     \\
            S2      & 0.63 (0.52)& 677 (830)          &  955 (1042)    &  15 ( 9)  &  61.3     \\
        \hline                                                         
        \multicolumn{6}{c}{B$_7$ (10.9~\%)}  \\                          
            S1      & 0.76 (0.70)& 635 (674)          &  880  (908)    &  23 (20)  &  63.1     \\
            S2      & 0.87 (0.55)& 454 (759)          &  801  (999)    &  22 (11)  &  69.4     \\
        \hline                                                         
        \multicolumn{6}{c}{B$_8$ (12.5~\%)}  \\                          
            S1$_a$  & 0.89 (0.70)& 509  (628)          &  798  (892)    &  26 (18)  &  60.0     \\
 S1$_b$ $\equiv$ S3 & 0.96 (0.89)& 606  (646)          &  841  (868)    &  35 (33)  &  80.7     \\
            S2      & 1.20 (0.75)& 372  (628)          &  718  (902)    &  31 (22)  &  71.3     \\
        \hline                                                         
        \multicolumn{6}{c}{B$_9$ (14.1~\%)}  \\                          
            S1$_a$  & 0.92 (0.78)& 577  (659)          &  825  (891)    &  31 (25)  &  58.6     \\
            S3$_a$  & 1.27 (0.84)& 414  (627)          &  708  (860)    &  37 (28)  &  70.1     \\
            S3$_b$  & 1.11 (0.80)& 494  (672)          &  777  (909)    &  37 (27)  &  60.3     \\
    \end{tabular}
\end{ruledtabular}
\end{table}

Despite their high density of states, low doping structures such as B$_4$C$_{60}$ and B$_5$C$_{59}$ 
cases do not exhibit strong electron-phonon interaction.
As we described in Fig.~\ref{fig:0dos}(c), the original electronic states around Fermi level comes from
$p$-orbital of 3-fold coordinated carbon atoms. 
This orbital character remains similar even when, for example, four boron atoms are doped.
The $\Gamma$ point wavefunction of the 128th state exhibits an almost localized feature around 3-fold coordinated atoms. 
The 127th state also has non-negligible amplitude around 3-fold coordinated atoms.
Even though the doping creates shallow acceptor levels, these states still do not contribute to the electron-phonon coupling. 
Therefore, it is important to increase the ``effective'' density of states that contributes to the electron-phonon coupling.

In fact, $\lambda$ slightly increases from 0.89 to 0.92 
although we put a boron atom in a 3-fold coordinated site of B$_8$-S1 structure to make the B$_9$-S1 structure. 
This doping creates a localized state around 1.5~eV as shown in the bottom 
panel of Fig.~\ref{fig:dos}, but the effective acceptor states does not decrease since
the Fermi energy is deep enough in the occupied states.
The formation of the 3-fold coordinated boron atom also modifies the phonon density of states 
as can be seen in the increase of the $\omega_{log}$ from 509 to 577~K.
Eventually the $T_c$ also increases from 26 to 31~K. 
This result interestingly suggests some 3-fold coordination 
should be allowed in a relatively high-density boron doping case.

\begin{figure}[thp]
  \includegraphics[width = 8.6cm]{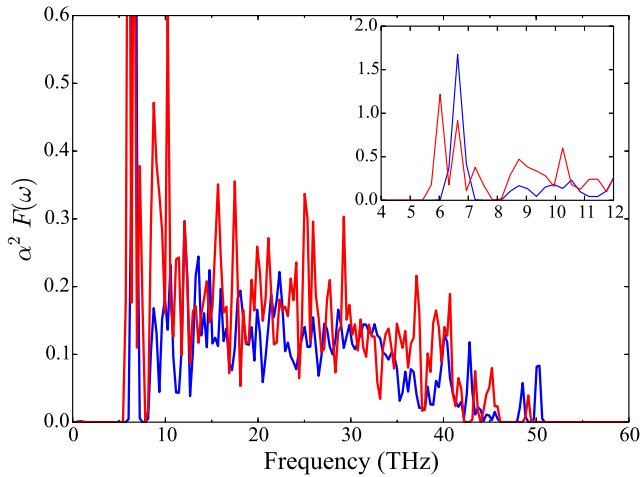}
    \caption{
    \label{fig:a2f}
(Color online) Eliashberg spectral function $\alpha^2F(\omega)$ of B$_4$-S1 (blue line) 
and B$_9$-S1$_a$ (red line) structures. The inset shows the magnified view of the low-frequency region. 
}
\end{figure}

Figure~\ref{fig:a2f} shows the Eliashberg spectral function $\alpha^2F(\omega)$ 
of B$_4$-S1 (blue line) and B$_9$-S1 (red line) structures. 
Comparing these two spectral functions, increasing  boron doping raises the 
spectral weight for almost the entire frequency range.
The spectral functions also show that spectral weight in low-frequency region is relatively large and irregular 
in the frequency region around 10~THz.
The low-frequency oscillations occur because we compute $\lambda$ at only the $\Gamma$ point.  Including
contributions from other $q$-points should decrease the large contribution from particular modes. 
In fact, when we compute the electron-phonon interaction of B$_8$-S1 case 
on $2\times2\times2$ $q$-grid and $4\times4\times4$ $k$-grid, 
the $\alpha^2F(\omega)$ becomes more regular and broadened, and the spectral weight in higher-frequency modes
become more substantial (see Fig.~S4 in Supplemental Material) \cite{SM}. 
The $\lambda$, $\omega_{log}$, and $T_c$ calculated from this spectral function are 0.80, 586~K, and 24~K, respectively.
This $\lambda$ is slightly smaller than the $\Gamma$ only calculation (0.89), but
$\omega_{log}$ becomes larger (509~K in the $\Gamma$ case), resulting in an almost similar value of $T_c$. 
Therefore,  $\Gamma$-only sampling should be an acceptable approximation particularly in estimating $T_c$. 

In Table~\ref{tb}, superconducting parameters without the contribution from the lowest-frequency optical phonon mode
are also listed. 
This is a rough approximation that reduces the spurious contribution from a particular low-frequency mode.
From the $2\times2\times2$ $q$-grid results on B$_8$-S1 case discussed above, 
these values are considered to be the worst-case scenario and the values computed 
with a sufficient number of $q$-points should be in between those with and without the lowest-frequency optical mode.
Nonetheless, the highest value of $T_c$ in this approximation is 33~K in 12.5~\% doped case and 
still close to the experimental $T_c$ of boron-doped Q-carbon.

We note that an imaginary frequency acoustic branch appears when we compute the phonon dispersion relation with 
a $2\times2\times2$ $q$-grid in the cases of 12.5~\% boron doped case (Fig.~S3 in Supplemental Material) \cite{SM}.
We attribute its origin to our artificially imposed cubic symmetry and to the restricted number of atoms.
A similar geometry might appear under the extreme experimental condition such as high temperature melting and quick quenching in Q-carbon. 

In summary, we have shown that a simulated model of boron-doped amorphous carbon, such as Q-carbon, 
is a superconductor when properly doped with shallow acceptor states.
Our highest $T_c$ of 37~K in 14~\% boron doped case obtained here is comparable to that observed in boron-doped Q-carbon.
Shallow acceptor states appear when doped boron atoms are 4-fold coordinated even when boron dimers are formed. 
On the other hand, 3-fold boron atoms are generally not favored since they create deep acceptor states, 
but interestingly can increase $T_c$ through an increase in $\omega_{log}$ in a relatively high doping case.
The procedure used here and findings should be useful for designing doped amorphous carbon
with interesting physical properties.

\begin{acknowledgments}
YS and JRC acknowledge support from the U.S.~Department of Energy (DoE) for work 
on nanostructures from grant DE-FG02-06ER46286, and on algorithms by a subaward from the 
Center for Computational Study of Excited-State Phenomena in Energy Materials at 
the Lawrence Berkeley National Laboratory, which is funded by the U.S.~Department of Energy, 
Office of Science, Basic Energy Sciences, Materials Sciences and Engineering 
Division under Contract No.~DE-AC02-05CH11231, as part of the Computational Materials Sciences 
Program. Computational resources are provided in part by the National Energy Research 
Scientific Computing Center (NERSC) and the Texas Advanced Computing Center (TACC). 
MLC acknowledges support from the 
National Science Foundation Grant No.~DMR-1508412 and from the Theory of Materials Program at the 
Lawrence Berkeley National Lab funded by the Director, Office of Science and Office of Basic 
Energy Sciences, Materials Sciences and Engineering Division, U.S.~Department of 
Energy under Contract No. DE-AC02-05CH11231. MLC acknowledges useful discussions with Professor Jay Narayan.
\end{acknowledgments}


%
\end{document}